\documentclass[12pt]{article}
\usepackage[utf8]{inputenc}

\usepackage{natbib}
\bibpunct{(}{)}{;}{a}{}{,}
\usepackage[pdftex]{graphicx}
\usepackage{hyperref}
\usepackage{setspace}
\usepackage[margin=1in]{geometry}
\usepackage{amsmath}
\usepackage{amsthm}
\usepackage{amssymb}
\usepackage{bm}

\newtheorem{proposition}{Proposition}

\newcommand{\argmax}{\mathop{\rm arg~max}\limits}

\title{Not Linearly Correlated, But Dependent:\\
A Family of Normal Mode Copulas}

\author{Kentaro Fukumoto\thanks{Professor, Department of Political Science, Gakushuin University. Mailing address:~1-5-1 Mejiro, Toshima, Tokyo 171-8588, Japan. Telephone:~+81-3-5904-9258. Fax:~+81-3-5992-1006. Email address:~\href{mailto:Kentaro.Fukumoto@gakushuin.ac.jp}{\texttt{Kentaro.Fukumoto@gakushuin.ac.jp}}. 
ORCiD: 0000-0003-3704-9054. 
Twitter: @000fukumoto.
}
}

\begin{document}

\maketitle

\begin{abstract}
When scholars study joint distributions of multiple variables, copulas are useful. However, if the variables are not linearly correlated with each other yet are still not independent, most of conventional copulas are
not up to the task. Examples include (inversed) U-shaped relationships and heteroskedasticity. To fill this gap, this manuscript sheds new light on a little-known copula, which I call the `normal mode copula.' I characterize the copula's properties and show that the copula is asymmetric and nonmonotonic under certain conditions. I also apply the copula to a dataset about U.S.~House vote share and campaign expenditure to demonstrate that the normal mode copula has better performance than other conventional copulas.

\vspace{0.2in}

\vspace{0.2in}
\end{abstract}

\section{Introduction}

It becomes hard to analyze multiple variables when they are generated at the same time. For example, it is not straightforward to regress one variable (as an outcome) on another (as a cause) without assuming the causal direction from the latter variable to the former. One approach is analyzing the joint distribution. Typically, scholars employ Heckman's sample selection models and the seemingly unrelated regression. However, a problem of these models is that they assume the bivariate normal distribution alone. 
By contrast, copula functions are able to represent a variety of dependence structures flexibly. (I will explain the copula shortly.) 

\begin{figure}[!htb]
\centering
\includegraphics[width = 5in]{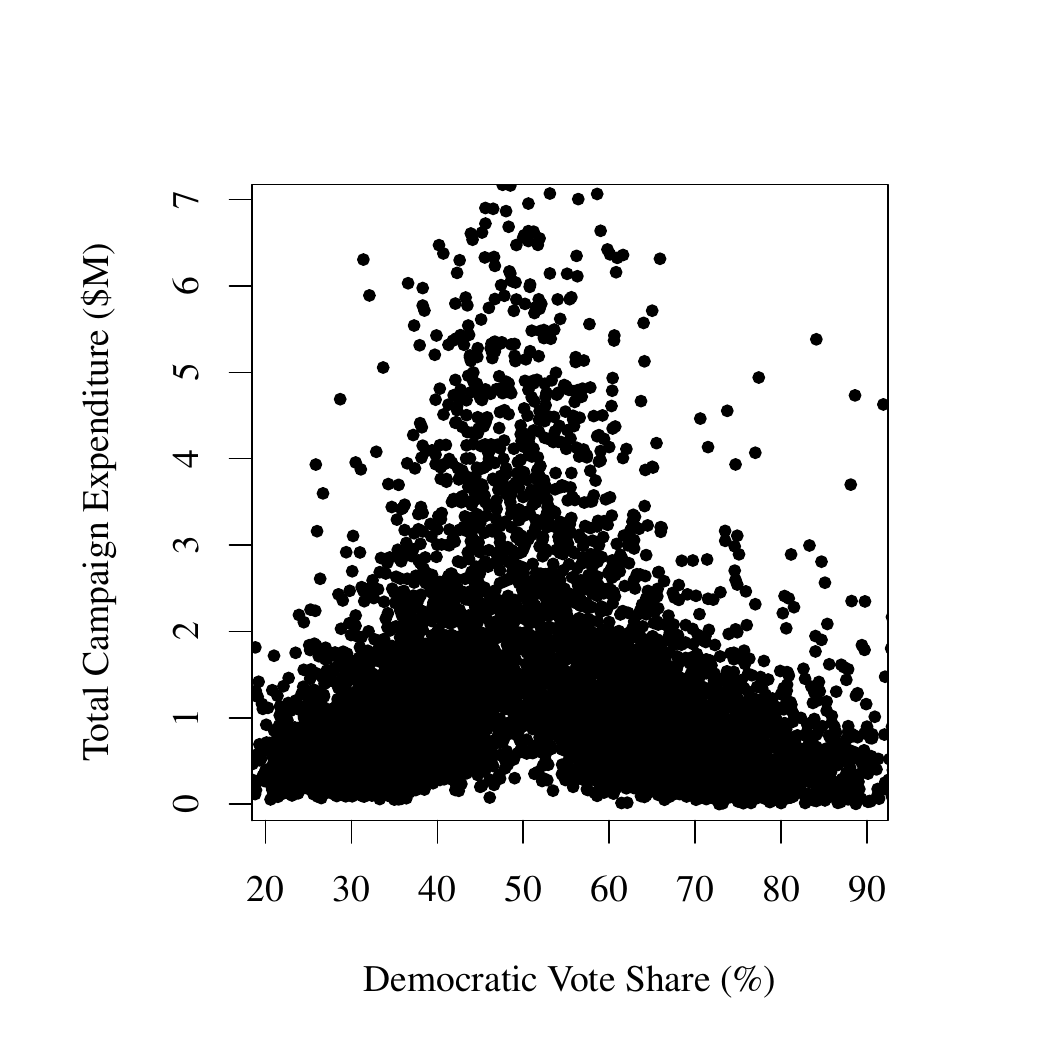}
\caption{Joint Distribution of Vote Share and Expenditure, U.S.~House Elections, 1946-2014}
\label{empirical_example}
\end{figure}

Nonetheless, if the variables are not linearly correlated with each other yet are still not independent, most of conventional copulas are not up to the task. 
Examples include (inversed) U-shaped relationships and heteroskedasticity. 
Figure \ref{empirical_example} shows a case of inversed U-shaped relationship. 
This is based on the data about U.S.~House elections, 1946--2014 \citep{Jacobson:2015}.
Each dot corresponds to a unit of observation, namely, a district in a year ($N=6,509$).
The horizontal axis represents the Democratic candidate's share of the two-party vote by percentage.
The vertical axis indicates the total amount of expenditure by both candidates (in million dollars).\footnote{I only include conventional units where one Democratic candidate and one Republican candidate ran and one of them was the incumbent (but not both). The amount of expenditure is adjusted for inflation ($2014 = 1.00$).
On both axes, I do not include extreme units where the value of either variable is smaller than the corresponding 1\% quantile value or larger than the corresponding 99\% quantile value.}
Here is a clear inverse U-shape.
When the race is competitive, both camps spend a lot, and the vote share is almost half and half. Otherwise, vote sharing is lopsided, and thus, neither candidate does not (have to) raise much campaign funding \citep{EriPal:00, JacKer:83}.
The Kendall's correlation coefficient between the two variables is $-0.076$. That is, the two variables have a linearly negative correlation significantly but only slightly. Obviously, the two are far from independent of each other.

\begin{center}
[Figure 1 about here]
\end{center}

In many situations like this case, underlying copulas are likely to be asymmetric and/or nonmonotonic, although conventional copulas are not. To fill this gap, I shed new light on a little-known copula with trigonometric functions, which I call the `normal mode copula' \citep{Fukumoto:2023}.

This manuscript is organized as follows. In the next section, I provide the definition of the normal mode copula, characterize its properties, and present statistical tools for copulas in general. The third section applies a normal mode copula to the above data about vote share and campaign expenditure to demonstrate that the normal mode copula models the data better than other conventional copulas. Finally, I offer discussion.

\section{Materials and Methods}

\subsection{Definition}

\subsubsection{Bivariate Copula}\label{sec:definition}

Suppose $0 \leq u_{d} \leq 1$ for $d \in \{1, 2\}$. A function $C(u_{1}, u_{2}): [0, 1]^2 \to [0, 1]$ is said to be a copula if it satisfies the following two conditions \citep[10]{Nelsen:06}:
\begin{description}
\item[Boundary conditions:] $C(u_{1}, 0) = C(0, u_{2}) = 0, C(u_{1}, 1) = u_{1}$, and $C(1, u_{2}) = u_{2}$.
\item[Two-increasing Condition:] If $u_{1}^{L} \leq u_{1}^{H}$ and $u_{2}^{L} \leq u_{2}^{H}$, it follows that $C(u_{1}^{H}, u_{2}^{H}) - C(u_{1}^{L}, u_{2}^{H}) - C(u_{1}^{H}, u_{2}^{L}) + C(u_{1}^{L}, u_{2}^{L}) \geq 0$.
\end{description}

Here is the motivation for copulas. 
For two random variables, $X_{1}$ and $X_{2}$, 
I denote the value of the $d$-th variable $X_{d}$ by $x_{d}$, the marginal cumulative distribution function (CDF) of $X_{d}$ by $F_{d}(x_{d}) \equiv u_{d}$, and the joint CDF of $X_{1}$ and $X_{2}$ by $F_{12}(x_{1}, x_{2})$.
Then, according to Sklar's theorem \citep[18 and 24--25]{Nelsen:06}, there is a copula $C$ such that
\begin{equation*}
F_{12}(x_{1}, x_{2}) = C \{ F_{1}(x_{1}), F_{2}(x_{2}) \}.
\end{equation*}
If $F_{1}$ and $F_{2}$ are continuous, $C$ is unique. 

There are dozens of copulas \citep{Hofert:2017, Nelsen:06, TriZim:07}. 
One of the well-known copulas is the Gaussian copula:
\begin{equation*}
C_{\mathrm{G}}(u_{1}, u_{2}) \equiv \Phi^2 \{ \Phi^{-1}(u_{1}), \Phi^{-1}(u_{2}) \mid \theta \},
\end{equation*}
where $\Phi$ and $\Phi^2$ are univariate and bivariate standard normal distributions, respectively, and $-1 \leq \theta \leq 1$ is the correlation parameter. Unless $F_{1}$ and $F_{2}$ are univariate normal distributions, $F_{12}$ is not a bivariate normal distribution. However, even if $F_{1}$ and $F_{2}$ are univariate normal distributions, $F_{12}$ is not a bivariate normal distribution unless $C$ is a Gaussian copula.

Now I introduce a family of copulas studied in this manuscript:
\begin{proposition}[Bivariate Normal Mode Copula]\label{prop:definition}
Let $\kappa_{1}$ and $\kappa_{2}$ be positive integers, $\theta$ be a real number, and
\begin{equation*}
C_{\mathrm{NM}} (u_{1}, u_{2}) \equiv u_{1} u_{2} + \frac{1}{\kappa_{1} \kappa_{2} \pi^2}\theta \sin (u_{1} \kappa_{1} \pi) \sin (u_{2} \kappa_{2} \pi).
\end{equation*}
Then, $C_{\mathrm{NM}}$ is a copula if and only if $ -1 \leq \theta \leq 1$.
\end{proposition}

Proof of all propositions are in the Supplementary Material. For $ -1 \leq \theta \leq 1$, I call $C_{\mathrm{NM}}$ the normal mode copula \citep{Fukumoto:2023}, $\theta$ the amplitude, and $\kappa_{1}$ and $\kappa_{2}$ mode numbers.

\begin{figure}[!htb]
\centering
\includegraphics[width = 5in]{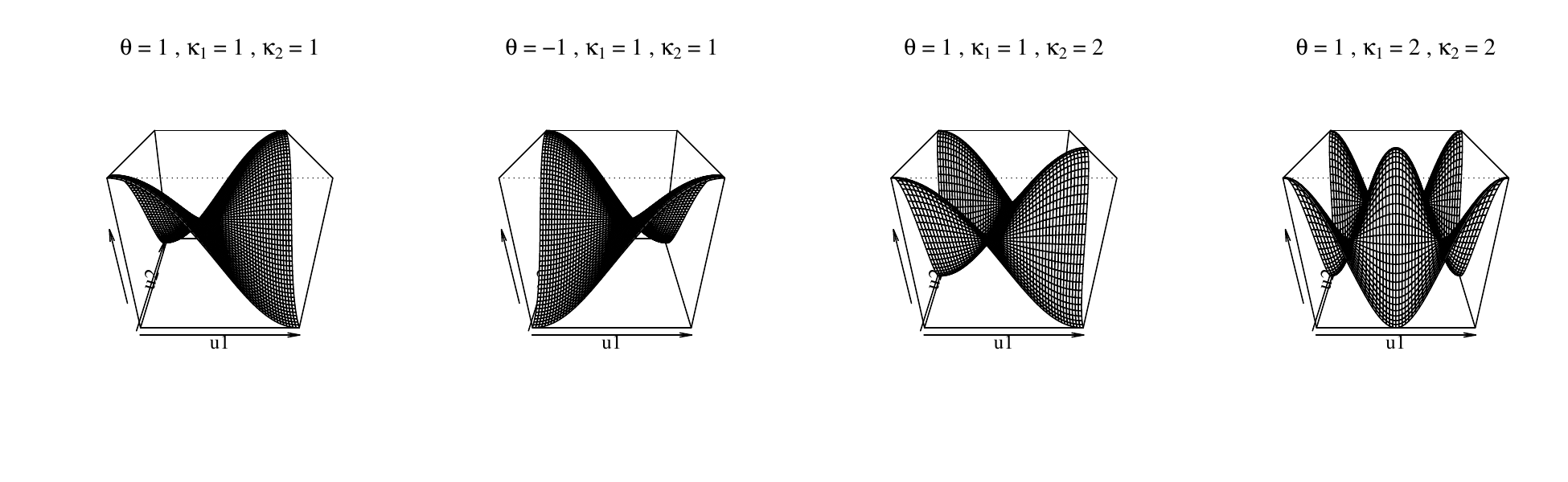}
\caption{Example Plots of Normal Mode Copulas}
\label{example.plot}
\end{figure}

Figure \ref{example.plot} shows three-dimensional plots of densities for normal mode copulas,
\begin{equation*}
\begin{split}
c_{\mathrm{NM}} (u_{1}, u_{2}) & \equiv \frac{\partial^2 }{\partial u_{1} \partial u_{2}} C_{\mathrm{NM}} (u_{1}, u_{2})\\
&= 1 + \theta \cos (u_{1} \kappa_{1} \pi) \cos (u_{2} \kappa_{2} \pi),
\end{split}
\end{equation*}
where the horizontal and vertical axes are $u_{1}$ and $u_{2}$, respectively. 
The first to fourth panels correspond to the cases where $(\theta, \kappa_{1}, \kappa_{2})$ is equal to $(1, 1, 1), (-1, 1, 1), (1, 1, 2)$ and $(1, 2, 2)$, respectively. 
In all panels except for the third, the normal mode copulas are (radially) symmetric. (I elaborate on the exact definition of copula's properties shortly.) 
The normal mode copulas in the first and second cases have a linearly positive and negative correlation, respectively, although those in the first and second cases have no linear correlation.
In the third panel, for $U_{2} \equiv F_{2}(X_{2}) \leq \frac{1}{2}$, $U_{1} \equiv F_{1}(X_{1})$ and $U_{2}$ have a linearly positive correlation, while they have a linearly negative correlation for $U_{2} \geq \frac{1}{2}$; thus, in sum, both correlation structures are canceled out, and $U_{1}$ and $U_{2}$ are not linearly correlated. However, the two variables are still not independent. Accordingly, the copula represents a typical situation of interest for this manuscript (cf., Figure \ref{empirical_example}).

\begin{center}
[Figure 2 about here]
\end{center}

\subsubsection{Multivariate Copula}\label{sec:definition_multivariate}

It is straightforward to extend the definition of a bivariate copula to its multivariate version.
Let $0 \leq u_{d} \leq 1$ for $d \in \{1, 2, \ldots, D \}$ and $\bm u = (u_{1}, \ldots, u_{D})$. A function $C(\bm u): [0, 1]^D \to [0, 1]$ that is continuous in $\bm u$ is called a copula if the following two conditions are met \citep[c.f.,][45]{Nelsen:06}:
\begin{description}
\item[Multivariate Boundary Conditions:] $C(\bm u) = 0$ if $u_{d}=0$ for some $d$ and $C(\bm u) = u_{d}$ if $u_{\tilde{d}} = 1$ for all $\tilde{d}$ such that $\tilde{d} \neq d$.
\item[$D$-increasing Condition:] $C(\bm u) $ is $D$ times differentiable and $c (\bm u) \equiv \frac{\partial^{D} }{\partial u_{1} \partial u_{2} \ldots \partial u_{D}} C (\bm u) \geq 0$.
\end{description}

I extend a bivariate normal mode copula to a multivariate normal mode copula as follows.
\begin{proposition}[Multivariate Normal Mode Copula]\label{Multivariate}
Let $\kappa_{1}, \kappa_{2}, \ldots, \kappa_{D}$ be positive integers for $d \in \{1, 2, \ldots, D \}$, and
\begin{equation*}
C_{\mathrm{NM}} (\bm u) \equiv \prod_{d = 1}^{D} u_{d} + \theta \prod_{d = 1}^{D} \frac{1}{\kappa_{d} \pi} \sin (u_{d} \kappa_{d} \pi) .
\end{equation*}
Then, $C_{\mathrm{NM}}$ is a copula if and only if $-1 \leq \theta \leq 1$.
\end{proposition}

A merit of the multivariate normal mode copula is that it is faster to calculate its value, $C_{\mathrm{NM}} (\bm u)$, and its density,
\begin{equation*}
c_{\mathrm{NM}} (\bm u) \equiv 1 + \theta \prod_{d = 1}^{D} \cos (u_{d} \kappa_{d} \pi),
\end{equation*}
than other multivariate copulas, for instance, the multivariate Gaussian copula.

\subsection{Properties}\label{sec:Properties}

Below, I focus on the bivariate normal mode copula and characterize its properties. For an introduction to copulas in general, refer to \citet{Nelsen:06}.

\subsubsection{Symmetricity}

In general, a copula $C$ is symmetric if $C (u_{1}, u_{2}) = C (u_{2}, u_{1})$ for all $u_{1}$ and $u_{2}$ (\citeauthor{Nelsen:06} \citeyear{Nelsen:06}, 38).

\begin{proposition}[Symmetricity]\label{Symmetricity}
A normal mode copula is symmetric if and only if $\kappa_{1} = \kappa_{2}$.
\end{proposition}

In general, for a copula, the following three copulas are called associated copulas \citep[13--14]{TriZim:07}:
\begin{equation*}
\begin{split}
\overline{C}^{(1)} (u_{1}, u_{2}) & \equiv u_{2} - C (1 - u_{1}, u_{2})\\
\overline{C}^{(2)} (u_{1}, u_{2}) & \equiv u_{1} - C (u_{1}, 1 - u_{2})\\
\overline{C}^{(12)} (u_{1}, u_{2}) & \equiv u_{1} + u_{2} - 1 + C (1 - u_{1}, 1 - u_{2}).
\end{split}
\end{equation*}
In particular, $\overline{C}^{(12)} (u_{1}, u_{2})$ is called the survival copula \citep[32]{Nelsen:06}.

\begin{proposition}[Associated Copula]\label{Associated.Copulas}
(1) For $d \in \{1,2\}$,
\begin{equation*}
\begin{split}
\overline{C}^{(d)}_{\mathrm{NM}} (\cdot, \cdot \mid \theta) =
\begin{cases} 
C_{\mathrm{NM}} (\cdot, \cdot \mid \theta) & \quad \textrm{if $\kappa_{d}$ is even} \\
C_{\mathrm{NM}} (\cdot, \cdot \mid - \theta)& \quad \textrm{otherwise.} 
\end{cases}
\end{split}
\end{equation*}
(2)
\begin{equation*}
\begin{split}
\overline{C}^{(12)}_{\mathrm{NM}} (\cdot, \cdot \mid \theta) =
\begin{cases} 
C_{\mathrm{NM}} (\cdot, \cdot \mid \theta) \quad &\textrm{if $\kappa_{1} + \kappa_{2}$ is even} \\
C_{\mathrm{NM}} (\cdot, \cdot \mid - \theta) \quad & \textrm{otherwise.} 
\end{cases}
\end{split}
\end{equation*}
\end{proposition}

Thus, unless either $\kappa_{1}$ or $\kappa_{2}$, but not both, are odd, a normal mode copula is radially symmetric ($\overline{C}^{(12)}  =C$) (\citeauthor{Nelsen:06} \citeyear{Nelsen:06}, 37). If $\kappa_{d}$ is even, a normal mode copula is symmetric in $u_{d}$ ($\overline{C}^{(d)} =C$).

\subsubsection{Dependence}

\noindent\underline{Monotonicity.} In general, I introduce six ways of defining monotonicity. I denote $X_{-d} \equiv X_{3 - d}$. (i) $X_{1}$ and $X_{2}$ are positive quadrant dependent (PQD($X_{1}, X_{2}$)) if $\Pr ( X_{1} \leq x_{1}, X_{2} \leq x_{2}) \geq \Pr (X_{1} \leq x_{1}) \Pr(X_{2} \leq x_{2})$. $X_{1}$ and $X_{2}$ are negative quadrant dependent (NQD($X_{1}, X_{2}$)) if $\Pr ( X_{1} \leq x_{1}, X_{2} \leq x_{2}) \leq \Pr (X_{1} \leq x_{1}) \Pr(X_{2} \leq x_{2})$ (\citeauthor{Nelsen:06} \citeyear{Nelsen:06}, 187).
(ii) $X_{d}$ is left tail decreasing (or increasing) in $X_{-d}$ (LTD ($X_{d} \mid X_{-d}$) (or LTI ($X_{d} \mid X_{-d}$))) if $\Pr(X_{d} \leq x_{d} \mid X_{-d} \leq x_{-d})$ is a nonincreasing (or nondecreasing) function of $x_{-d}$ for all $x_{d}$ (\citeauthor{Nelsen:06} \citeyear{Nelsen:06}, 191).
(iii) $X_{d}$ is right tail increasing (or decreasing) in $X_{-d}$ (RTI($X_{d} \mid X_{-d}$) (or RTD($X_{d} \mid X_{-d}$))) if $\Pr(X_{d} > x_{d} \mid X_{-d} > x_{-d})$ is a nondecreasing (or nonincreasing) function of $x_{-d}$ for all $x_{d}$ (\citeauthor{Nelsen:06} \citeyear{Nelsen:06}, 191).
(iv) $X_{d}$ is stochastically increasing (or decreasing) in $X_{-d}$ (SI($X_{d} \mid X_{-d}$) (or SD($X_{d} \mid X_{-d}$))) if $\Pr(X_{d} \leq x_{d} \mid X_{-d} = x_{-d})$ is a nonincreasing (or nondecreasing) function of $x_{-d}$ for all $x_{d}$ (\citeauthor{Nelsen:06} \citeyear{Nelsen:06}, 196).
(v) $X_{1}$ and $X_{2}$ are left corner set decreasing (or increasing) (LCSD($X_{1}, X_{2}$) (or LCSI($X_{1}, X_{2}$))) if $\Pr ( X_{1} \leq x_{1}, X_{2} \leq x_{2} \mid X_{1} \leq \tilde{x}_{1}, X_{2} \leq \tilde{x}_{2})$ is nonincreasing (or nondecreasing) in $\tilde{x}_{1}$ and $\tilde{x}_{2}$ for all $x_{1}$ and $x_{2}$ (\citeauthor{Nelsen:06} \citeyear{Nelsen:06}, 198).
(vi) $X_{1}$ and $X_{2}$ are right corner set increasing (or decreasing) (RCSI($X_{1}, X_{2}$) (or RCSD($X_{1}, X_{2}$))) if $\Pr ( X_{1} > x_{1}, X_{2} > x_{2} \mid X_{1} > \tilde{x}_{1}, X_{2} > \tilde{x}_{2})$ is nondecreasing (or nonincreasing) in $\tilde{x}_{1}$ and $\tilde{x}_{2}$ for all $x_{1}$ and $x_{2}$ (\citeauthor{Nelsen:06} \citeyear{Nelsen:06}, 198).

\begin{proposition}[Monotonicity]\label{monotonicity} Suppose the copula of $X_{1}$ and $X_{2}$ is a normal mode copula.
\begin{description}
\item[(1) Positive Dependence:] For $d \in \{1, 2\}$, if and only if $\kappa_{1} = \kappa_{2} = 1, \theta > 0$ or $\theta = 0$, (i) PQD($X_{1}, X_{2}$), (ii) LTD($X_{d} \mid X_{-d}$), (iii) RTI($X_{d} \mid X_{-d}$), (iv) SI($X_{d} \mid X_{-d}$), (v) LCSD($X_{1}, X_{2}$), and (vi) RCSI($X_{1}, X_{2}$).
\item[(2) Negative Dependence:] For $d \in \{1, 2\}$, if and only if $\kappa_{1} = \kappa_{2} = 1$ and $\theta < 0$ or $\theta = 0$, (i) NQD($X_{1}, X_{2}$), (ii) LTI($X_{d} \mid X_{-d}$), (iii) RTD($X_{d} \mid X_{-d}$), (iv) SD($X_{d} \mid X_{-d}$), (v) LCSI($X_{1}, X_{2}$), and (vi) RCSD($X_{1}, X_{2}$).
\end{description}
\end{proposition}

Therefore, unless $\kappa_{1} = \kappa_{2} = 1$ or $\theta = 0$, the dependence between $X_{1}$ and $X_{2}$ is nonmonotonic.

In general, if $C^{(1)}(u_{1}, u_{2}) \leq C^{(2)} (u_{1}, u_{2})$ for all $u_{1}$ and $u_{2}$, $C^{(2)}$ is said to be more PQD (or concordant) than $C^{(1)}$, and I note that $C^{(1)} \prec_{\mathrm{C}} C^{(2)}$. A family of copulas $C (\cdot, \cdot \mid \theta)$ indexed by $\theta$ is ordered in concordance if $\theta^{L} \leq \theta^{H}$ implies $C (\cdot, \cdot \mid \theta^{L}) \prec_{\mathrm{C}} C (\cdot, \cdot \mid \theta^{H})$ \citep[39]{Amblard_Girard_2002, Nelsen:06}.

It holds that \citep[41]{Nelsen:06}
\begin{equation*}
\begin{split}
\Pr(X_{d} \leq x_{d} \mid X_{-d} = x_{-d})
& = \frac{\partial  }{\partial u_{-d}} C (u_{1}, u_{2})\\
& \equiv C^{\prime}_{d \mid -d} ( u_{d} \mid u_{-d}), 
\end{split}
\end{equation*}
where $u_{-d} \equiv u_{3 - d}$.
If $C_{2 \mid 1}^{(2) \prime -1} \{ C_{2 \mid 1}^{(1) \prime} ( u_{2} \mid u_{1}) \mid u_{1} \}$ is nondecreasing in $u_{1}$, $C^{(2)}$ is said to be more SI than $C^{(1)}$, and I note that $C^{(1)} \prec_{\mathrm{SI}} C^{(2)}$. A family of copulas $C (\cdot, \cdot \mid \theta)$ indexed by $\theta$ is SI ordered if $\theta^{L} \leq \theta^{H}$ implies $C (\cdot, \cdot \mid \theta^{L}) \prec_{\mathrm{SI}} C (\cdot, \cdot \mid \theta^{H})$ \citep{Amblard_Girard_2002}.

\begin{proposition}[Ordering within Normal Mode Copulas]\label{Dependence.Ordering}
(1) The family of normal mode copulas is ordered in concordance if and only if $\kappa_{1} = \kappa_{2} = 1$.

(2) The family of normal mode copulas is SI ordered if and only if $\kappa_{1} = \kappa_{2} = 1$.
\end{proposition}

\noindent\underline{Tail Dependence.} In general, the lower and upper tail dependence parameters are defined as
\begin{equation*}
\begin{split}
\lambda_{\mathrm{L}} & \equiv \lim_{u \to 0^{+}} \Pr \{ X_{2} \leq F_{2}^{-1} (u) \mid X_{1} \leq F_{1}^{-1} (u) \}\\
\lambda_{\mathrm{H}} & \equiv \lim_{u \to 1^{-}} \Pr \{ X_{2} > F_{2}^{-1} (u) \mid X_{1} > F_{1}^{-1} (u) \},
\end{split}
\end{equation*}
respectively, if they exist \citep[214]{Nelsen:06}. In some disciplines, such as finance, reliance analysis, and disaster studies, scholars are interested in how often one variable goes extreme when the other does so.

\begin{proposition}[Tail Dependence]\label{Tail.Dependence}
Suppose the copula of $X_{1}$ and $X_{2}$ is a normal mode copula. Then, $\lambda_{\mathrm{L}} = \lambda_{\mathrm{H}} =0$.
\end{proposition}

This proposition implies that even if $X_{d}$ takes an extremely large or small value, $X_{-d}$ does not take an extreme value. In this case, $X_{1}$ and $X_{2}$ are said to be lower and upper tail independent.

\noindent\underline{Measures of Association.} In general, I introduce six measures of association that scale the degree to which $X_{1}$ and $X_{2}$ are dependent on each other.
I suppose that
\begin{equation*}
\begin{split}
X_{1}, X^{(1)}_{1}, X^{(2)}_{1} & \overset{\mathrm{iid}}{\sim} F_{1}(x_{1}) \\
X_{2}, X^{(1)}_{2}, X^{(2)}_{2} & \overset{\mathrm{iid}}{\sim} F_{2}(x_{2}) \\
(X_{1}, X_{2}) &\sim C \{ F_{1}(x_{1}), F_{2}(x_{2}) \} \\
(X^{(1)}_{1}, X^{(1)}_{2}) & \sim C^{(1)} \{ F_{1}(x^{(1)}_{1}), F_{2}(x^{(1)}_{2}) \} \\
(X^{(2)}_{1}, X^{(2)}_{2}) & \sim C^{(2)} \{ F_{1}(x^{(2)}_{1}), F_{2}(x^{(2)}_{2}) \}.
\end{split}
\end{equation*}
I denote
\begin{equation*}
\Omega (C^{(1)}, C^{(2)} ) \equiv \Pr \{ (X_{1}^{(1)} - X_{1}^{(2)}) (X_{2}^{(1)} - X_{2}^{(2)}) > 0\} - \Pr \{ (X_{1}^{(1)} - X_{1}^{(2)}) (X_{2}^{(1)} - X_{2}^{(2)}) < 0\},
\end{equation*}
where the first and second terms correspond to the probabilities of concordance and disconcordance, respectively \citep[159]{Nelsen:06}.

Kendall's $\tau$ for $X_{1}$ and $X_{2}$ is defined as (\citeauthor{Nelsen:06} \citeyear{Nelsen:06}, 161 and 164)
\begin{equation*}
\begin{split}
\tau & \equiv \Omega (C, C) \\
&=1 - 4 \int_0^1 \int_0^1 \frac{\partial }{\partial u_{1}}C(u_{1},u_{2})
\frac{\partial }{\partial u_{2}}C(u_{1},u_{2}) d u_{1} d u_{2}.
\end{split}
\end{equation*}

The product (or independence) copula is denoted by $C_{\mathrm{I}} (u_{1}, u_{2}) \equiv u_{1} u_{2}$, where $X_{1}$ and $X_{2}$ are independent of each other (\citeauthor{Nelsen:06} \citeyear{Nelsen:06}, 25).
Spearman's $\rho$ for $X_{1}$ and $X_{2}$ is defined as (\citeauthor{Nelsen:06} \citeyear{Nelsen:06}, 167 and 170)
\begin{equation*}
\begin{split}
\rho & \equiv \Omega (C, C_{\mathrm{I}})\\
&= 12 \int_0^1 \int_0^1 \{ C(u_{1}, u_{2}) - C_{\mathrm{I}} (u_{1}, u_{2}) \} 
d u_{1} d u_{2}.
\end{split}
\end{equation*}
By replacing the term inside integration by its absolute value, we obtain Schweizer and Wolff's $\sigma$ (or the normalized volume) for $X_{1}$ and $X_{2}$ (\citeauthor{Nelsen:06} \citeyear{Nelsen:06}, Nelsen:06),
\begin{equation*}
\sigma \equiv 12 \int_0^1 \int_0^1 | C(u_{1},u_{2}) - C_{\mathrm{I}} (u_{1}, u_{2}) | d u_{1} d u_{2}.
\end{equation*}

Fr\'echet lower and upper bound copulas are defined as
\begin{equation*}
\begin{split}
C_{\mathrm{L}}(u_{1}, u_{2}) &\equiv \max(u_{1} + u_{2} - 1, 0)\\
C_{\mathrm{H}}(u_{1}, u_{2}) &\equiv \min(u_{1}, u_{2}),
\end{split}
\end{equation*}
respectively. For any copula $C$, it holds that $C_{\mathrm{L}} \prec_{\mathrm{C}} C \prec_{\mathrm{C}} C_{\mathrm{H}}$ (\citeauthor{Nelsen:06} \citeyear{Nelsen:06}, 11). Gini's $\gamma$ for $X_{1}$ and $X_{2}$ is defined as (\citeauthor{Nelsen:06} \citeyear{Nelsen:06}, 180 and 181)
\begin{equation*}
\begin{split}
\gamma &\equiv \Omega (C, C_{\mathrm{L}}) + \Omega (C, C_{\mathrm{H}})\\
&= 4 \left[ \int^{1}_{0} C(u, 1 - u) du - \int^{1}_{0} \{ u - C(u, u) \} du \right].
\end{split}
\end{equation*}
Spearman's foot-rule $\phi$ for $X_{1}$ and $X_{2}$ is defined as (\citeauthor{Nelsen:06} \citeyear{Nelsen:06}, 186)
\begin{equation*}
\begin{split}
\phi & \equiv \frac{1}{2} \{ 3 \Omega (C, C_{\mathrm{H}}) - 1 \} \\
&= 6 \int^{1}_{0} C(u, u) du - 2.
\end{split}
\end{equation*}

Blomqvist's $\beta$ (or the medial correlation coefficient) for $X_{1}$ and $X_{2}$ is defined as (\citeauthor{Nelsen:06} \citeyear{Nelsen:06}, 182)
\begin{equation*}
\begin{split}
\beta &\equiv Pr \left[ \left\{ X_{1} - F_{1}^{-1}\left( \frac{1}{2} \right) \right\} \left\{ X_{2} -F_{2}^{-1}\left( \frac{1}{2} \right) \right\} > 0 \right] \\
& \quad - \Pr \left[ \left\{ X_{1} - F_{1}^{-1}\left( \frac{1}{2} \right) \right\} \left\{ X_{2} -F_{2}^{-1}\left( \frac{1}{2} \right) \right\} < 0 \right] \\
&=4 C \left(\frac{1}{2}, \frac{1}{2}\right) - 1.
\end{split}
\end{equation*}

Since all information of the dependence between $X_{1}$ and $X_{2}$ is concentrated on $C$, these measures of association are functions of $C$ and have nothing to do with marginal distributions, $F_{1}$ and $F_{2}$.

\begin{proposition}[Measures of Association]\label{Measures.of.Association} Suppose the copula of $X_{1}$ and $X_{2}$ is a normal mode copula. For $X_{1}$ and $X_{2}$,

(1) Schweizer and Wolff's $\sigma$ is equal to $\frac{48 }{\kappa_{1} \kappa_{2} \pi^4}|\theta|$.

(2) Spearman's $\rho$ is equal to $\frac{48 }{\kappa_{1}^2 \kappa_{2}^2 \pi^4}\theta$ if both $\kappa_{1}$ and $\kappa_{2}$ are odd and 0 otherwise.

(3) Kendall's $\tau$ is equal to $\frac{32 }{\kappa_{1}^2 \kappa_{2}^2 \pi^4}\theta$ if both $\kappa_{1}$ and $\kappa_{2}$ are odd and 0 otherwise.

(4) Blomqvist's $\beta$ is equal to
\begin{equation*}
\begin{split}
\beta &= 
\begin{cases} 
\frac{1 }{\kappa_{1} \kappa_{2} \pi^2}\theta &\quad \textrm{if $\kappa_{1} \bmod 4 = 1, \kappa_{2} \bmod 4 = 1$ or $\kappa_{1} \bmod 4 = 3, \kappa_{2} \bmod 4 = 3$}\\
- \frac{1}{\kappa_{1} \kappa_{2} \pi^2}\theta  &\quad \textrm{if $\kappa_{1} \bmod 4 = 1, \kappa_{2} \bmod 4 = 3$ or $\kappa_{1} \bmod 4 = 3, \kappa_{2} \bmod 4 = 1$}\\
0& \quad \textrm{otherwise.} 
\end{cases}
\end{split}
\end{equation*}

(5) Gini's $\gamma$ is equal to zero.

(6) Spearman's foot-rule $\phi$ is equal to zero.
\end{proposition}

For instance, when $\kappa_{1} = \kappa_{2} =1$, $\rho \approx 0.49 \theta$, which is modest.
If both $\kappa_{1}$ and $\kappa_{2}$ are odd, then
\begin{equation*}
\frac{1}{|\rho|}\sigma = \kappa_{1} \kappa_{2}, \quad
\frac{1}{\rho}\tau =\frac{2}{3},\quad
\left| \frac{1}{\rho}\beta \right| = \frac{1}{48}\kappa_{1} \kappa_{2} \pi^2.
\end{equation*}
In addition, the sizes of the four measures of association ($\sigma, |\rho|, |\tau|$, and $|\beta|$) decrease in $\kappa_{1}$ and $\kappa_{2}$.
Otherwise, importantly, $\tau$, $\rho$, and $\beta$ are equal to zero, even when $X_{1}$ and $X_{2}$ are not independent of each other ($\theta \neq 0$).

\subsubsection{Relation with Other Copulas}\label{sec:other}

\noindent\underline{Product Copula.} When the copula of $X_{1}$ and $X_{2}$ is $C_{\mathrm{I}}$, $X_{1}$ and $X_{2}$ are independent of each other \citep[25]{Nelsen:06}. This is because $C^{\prime}_{\mathrm{I}, d \mid -d} ( u_{d} \mid u_{-d}) = u_{d}$.

\begin{proposition}[Independence]\label{Independence}
$C_{\mathrm{NM}} = C_{\mathrm{I}}$ if and only if $\theta = 0$.
\end{proposition}

Thus, the normal mode copula contains the product copula as a special case.

\noindent\underline{Fr\'echet Bound Copulas.} I repeat that for any copula $C$, $C_{\mathrm{L}} \prec_{\mathrm{C}} C \prec_{\mathrm{C}} C_{\mathrm{H}}$.

\begin{proposition}[Bounds]\label{Bounds}
There are no $\theta, \kappa_{1}$, and $\kappa_{2}$ such that

(1) $C_{\mathrm{NM}} (\cdot, \cdot \mid \theta, \kappa_{1}, \kappa_{2}) = C_{\mathrm{L}}(\cdot, \cdot)$

(2) $C_{\mathrm{NM}} (\cdot, \cdot \mid \theta, \kappa_{1}, \kappa_{2}) = C_{\mathrm{H}}(\cdot, \cdot)$.
\end{proposition}

This proposition implies that Fr\'echet lower and upper bound copulas do not belong to the family of normal mode copulas. In this sense, the family is said not to be comprehensive; that is, the degree of association between $X_{1}$ and $X_{2}$ does not reach the permissible range (as Proposition \ref{Measures.of.Association} suggests).

\noindent\underline{FGM Copulas.}\label{FGM} 
The Farlie-Gumbel-Morgenstein (FGM) copula is defined as (\citeauthor{Nelsen:06} \citeyear{Nelsen:06}, 77)
\begin{equation*}
C_{\mathrm{FGM}} (u_{1}, u_{2}) \equiv u_{1} u_{2} + \theta u_{1} (1 - u_{1}) u_{2} (1 - u_{2}),
\end{equation*}
where $-1 \leq \theta \leq 1$.

\begin{proposition}[Ordering with FGM Copulas]\label{Ordering}
Define
\begin{description}
\item[Condition (i):] $\kappa_{1} \bmod 4 = \kappa_{2} \bmod 4 =1$ or $\kappa_{1} \bmod 4 = \kappa_{2} \bmod 4 = 3$
\item[Condition (ii):] $\kappa_{1} \bmod 4 = 1, \kappa_{2} \bmod 4 =3$ or $\kappa_{1} \bmod 4 = 3, \kappa_{2} \bmod 4 = 1$.
\end{description}
Denote
\begin{equation*}
\theta^{*} \equiv \frac{1 }{4}| \theta | \kappa_{1} \kappa_{2} \pi^2.
\end{equation*}
If Condition (i), $\theta \geq 0, \theta_{\mathrm{FGM}} \geq \theta^{*}$ or Condition (ii), $\theta \leq 0, \theta_{\mathrm{FGM}} \geq \theta^{*}$,
\begin{equation*}
C_{\mathrm{NM}} (\cdot, \cdot \mid \theta) \prec_{\mathrm{C}} C_{\mathrm{FGM}} (\cdot, \cdot \mid \theta_{\mathrm{FGM}}).
\end{equation*}
If Condition (i), $\theta \leq 0, \theta_{\mathrm{FGM}} \leq -\theta^{*}$ or Condition (ii), $\theta \geq 0, \theta_{\mathrm{FGM}} \leq -\theta^{*}$,
\begin{equation*}
C_{\mathrm{NM}} (\cdot, \cdot \mid \theta) \succ_{\mathrm{C}} C_{\mathrm{FGM}} (\cdot, \cdot \mid \theta_{\mathrm{FGM}}).
\end{equation*}
\end{proposition}

This proposition implies that an FGM copula can be either more PQD or NQD than some normal mode copulas when the parameter of the FGM copula is sufficiently large or small.

\noindent\underline{Archimedean Copulas.} 
Let generator function $\varphi(u): [0, 1] \to [0, \infty]$ be a continuous, convex, strictly decreasing function, where $\varphi(1) = 0$. I define the pseudoinverse function of $\varphi(u)$ as $\varphi^{[-1]}(z) = \varphi^{-1}(z)$ for $0 \leq z \leq \varphi(0)$ and $\varphi^{[-1]}(z) = 0$ for $\varphi(0) \leq z \leq \infty$. Then, the following function
\begin{equation*}
C_{\mathrm{A}} ( u_{1}, u_{2} \mid \varphi) \equiv \varphi^{[-1]}\{ \varphi(u_{1}) + \varphi(u_{2}) \}
\end{equation*}
meets the two conditions of a copula and is called an Archimedean copula \citep[110--112]{Nelsen:06}. Archimedean copulas are one of the most studied, large classes of copulas.

\begin{proposition}[Archimedean]\label{Archimedean}
There are no $\theta \neq 0, \kappa_{1}, \kappa_{2}$, and $\varphi$ such that $C_{\mathrm{NM}} (\cdot, \cdot \mid \theta, \kappa_{1}, \kappa_{2})= C_{\mathrm{A}} (\cdot, \cdot \mid \varphi)$.
\end{proposition}

This proposition implies that a normal mode copula is not an Archimedean copula except when $\theta = 0$.

\noindent\underline{Circular Copulas.} 
I suppose $X_{1}$ is a circular variable (e.g., direction, the date of the year) such that its conditional `density,' $f_{1 \mid 2} (x_{1} \mid x_{2})$, is continuous in $x_{1}$,
$f_{1 \mid 2} (x_{1}  + 2 \pi \mid x_{2}) = f_{1 \mid 2} (x_{1} \mid x_{2})$, and $\int^{2 \pi}_{0} f_{1 \mid 2} (x_{1} \mid x_{2}) dx_{1} = 1$ \citep{Gill_Hangartner_2010}. Only in this underline, I redefine $F_{1} (x_{1}) \equiv \int^{x_{1}^{*}}_{0} f_{1} (x) dx$, where $x_{1} \bmod 2 \pi = x_{1}^{*}$ and $0 \leq x_{1}^{*} < 2 \pi$.
Since $f_{1 \mid 2} (x_{1} \mid x_{2}) = c \{ F_{1} (x_{1}), F_{2} (x_{2}) \} f_{1} (x_{1}), f_{1 \mid 2} (0 \mid x_{2}) = f_{1 \mid 2} (2 \pi \mid x_{2})$, and thus $f_{1} (0) = f_{1} (2 \pi)$, it follows that
\begin{equation*}
c (0, u_{2}) = \lim_{u_{1} \to 1} c (u_{1}, u_{2}).
\end{equation*}
\citet{Hodel_Fieberg_2022} call this type of copula a circular-linear copula. Most of conventional copulas do not satisfy this condition. In the case of normal mode copulas, this condition holds if and only if $\kappa_{1}$ is even. Furthermore, if $X_{2}$ is also a circular variable (where the corresponding copula is a circular-circular copula, or `circulas' according to \citet{Jones_Pewsey_Kato_2015}), $\kappa_{2}$ should be even as well.

\noindent\underline{Other Copulas with Trigonometric Functions.} 
Recently, scholars have proposed many copulas with trigonometric functions. I review them in the Supplementary Material.

\citet{Chesneau_2021_AM} refers to an extended version of the normal mode copula but only in passing,
\begin{equation*}
C_{\mathrm{ENM}} (\bm u) = \prod_{d = 1}^{D} u_{d} + \theta \prod_{d = 1}^{D} \frac{ 1 }{ p_{d} \kappa_{d} \pi }\{ \sin (u_{d} \kappa_{d} \pi ) \}^{p_{d}},
\end{equation*}
where $p_{d} \geq 1$ and $-1 \leq \theta \leq 1$.
This class of copulas includes the multivariate normal mode copulas if $p_{d} = 1$ for all $d$.
Nonetheless, \citet{Chesneau_2021_AM} mentions that `[t]his copula \textit{can} be asymmetric, orthant-dependent (positively or negatively) [i.e., PQD or NQD], radially symmetric, and with no tail dependence' (emphasis added) but neither considers the conditions of these properties nor examines other properties (such as monotonicity, measures of association, and relation with other copulas) in detail, nor applies it to any real data, leaving a comment, `this generalization remains unexplored.'

Other proposed copulas are different from the normal mode copulas unless a special case of them is the same as the normal mode copulas with $\kappa_{1} = \kappa_{2} ( = 1)$, whose properties are distinct from normal mode copulas with other values of $\kappa_{1}$ and $\kappa_{2}$ as detailed above. Most of the other proposed copulas cannot model the target of this manuscript, that is, variables that are not linearly correlated with, but still dependent on, each other. Some proposed copulas are Archimedean copulas, although the normal mode copulas are not (Proposition \ref{Archimedean}).

\subsection{Estimation and Evaluation}

This section briefly overviews methods of estimation and evaluation for copulas that I use in the following section, referring to \citet{Hofert:2017}.

\subsubsection{Maximum Pseudolikelihood Estimator}

I suppose there are $N$ units.
For $i \in \{1, 2, \ldots, N\}$ and $d \in \{1, 2\}$, unit $i$'s value of the $d$-th variable $X_{d}$ is denoted by $x_{d, i}$.

I also suppose we are agnostic about $F_{d}(x)$.\footnote{Instead, if we assume a parametric model of $F_{d}$ (sometimes with covariates), we may employ the maximum (nonpseudo) likelihood estimator or the inference functions for margins estimator  \citep[134--138]{Hofert:2017}.}
Then, a consistent estimate of $u_{d, i} \equiv F_{d}(x_{d, i})$ is its empirical quantile value (\citeauthor{Hofert:2017} \citeyear{Hofert:2017}, 139 and 199):
\begin{equation*}
\hat{u}_{d, i} \equiv \frac{1}{N + 1} \sum^{N}_{j = 1} \frac{1}{2} \{ I(x_{d, j} \leq x_{d, i}) + I(x_{d, j} < x_{d, i}) + 1 \},
\end{equation*}
where the summand is unit $i$'s rank and $I(\cdot)$ denotes the indicator function.\footnote{I handle tied units by using their average rank. I add one and divide the summation by $N+1$ rather than $N$ so that $0 < \hat{u}_{d, i} < 1$.}
We call $(\hat{u}_{1, i}, \hat{u}_{2, i})$ pseudoobservation.

I assume that $(U_{1, i}, U_{2, i})$ follows $C ( u_{1, i}, u_{2, i} \mid \theta )$.
I estimate $\theta$ by finding $\theta$ that maximizes the pseudolog-likelihood \citep[145]{Hofert:2017}:
\begin{equation*}
\hat{\theta} \equiv \argmax_{\theta} \sum^{N}_{i = 1} \log \{ c ( \hat{u}_{1, i}, \hat{u}_{2, i} \mid \theta ) \}.
\end{equation*}

\subsubsection{Goodness-of-fit: Cram\'er-von Mises Criterion}

For the goodness-of-fit of $C(\cdot, \cdot \mid \hat{\theta})$ to the data, we see the Cram\'er-von Mises criterion (CvMC) (\citeauthor{Hofert:2017} \citeyear{Hofert:2017}, 182):
\begin{equation*}
\textrm{CvMC} \equiv \sum^{N}_{i = 1} \{ C(\hat{u}_{1, i}, \hat{u}_{2, i} \mid \hat{\theta}) - \hat{C}(\hat{u}_{1, i}, \hat{u}_{2, i} ) \}^2,
\end{equation*}
where the summand is the square of the distance between the estimated copula and the empirical copula \citep[158]{Hofert:2017},
\begin{equation*}
\hat{C}(u_{1}, u_{2} ) \equiv \frac{1}{N} \sum^{N}_{j = 1} I(\hat{u}_{1, j} \leq u_{1}) I(\hat{u}_{2, j} \leq u_{2}),
\end{equation*}
evaluated at every pseudoobservation.

\subsubsection{Model Selection: Cross-validation Copula Information Criterion}

When scholars compare models, they usually pay attention to the Akaike information criterion (AIC). However, since we estimate $(u_{1, i}, u_{2, i})$ nonparametrically, and thus our estimation is not fully parametric, AIC is not necessarily the best option. Thus, we also employ the cross-validation copula information criterion (CIC), which is a natural adaptation of leave-one-out cross validation (\citeauthor{Hofert:2017} \citeyear{Hofert:2017}, 192),
\begin{equation*}
\textrm{CIC} \equiv \frac{1}{N} \sum^{N}_{i = 1} \log \{ c( \hat{u}_{1, -i}, \hat{u}_{2, -i} \mid \hat{\theta}_{-i}) \},
\end{equation*}
where for all $d$,
\begin{equation*}
\begin{split}
\hat{u}_{d, -i} & \equiv
\begin{cases} 
\frac{1}{N} \sum_{j \in \mathcal{J}_{-i}} I(\hat{u}_{d, j} \leq \hat{u}_{d, i}) &\quad \textrm{if $\hat{u}_{d, i} \geq \min_{j \in \mathcal{J}_{-i}} \hat{u}_{d, j}$}\\
\frac{1}{N} & \quad \textrm{otherwise} 
\end{cases}\\
\mathcal{J}_{-i} &\equiv\{1, 2, \ldots, N \} \setminus \{ i \},
\end{split}
\end{equation*}
and $\hat{\theta}_{-i}$ is the maximum pseudolikelihood estimate computed from the sample except for $i$'s unit:
\begin{equation*}
\hat{\theta}_{-i} \equiv \argmax_{\theta} \sum_{ j \in \mathcal{J}_{-i}} \log \{ c ( \hat{u}_{1, j}, \hat{u}_{2, j} \mid \theta ) \}.
\end{equation*}
The larger CIC is, the better.

\section{Results}\label{sec:Application}

\begin{figure}[!htb]
\centering
\includegraphics[width = 2in]{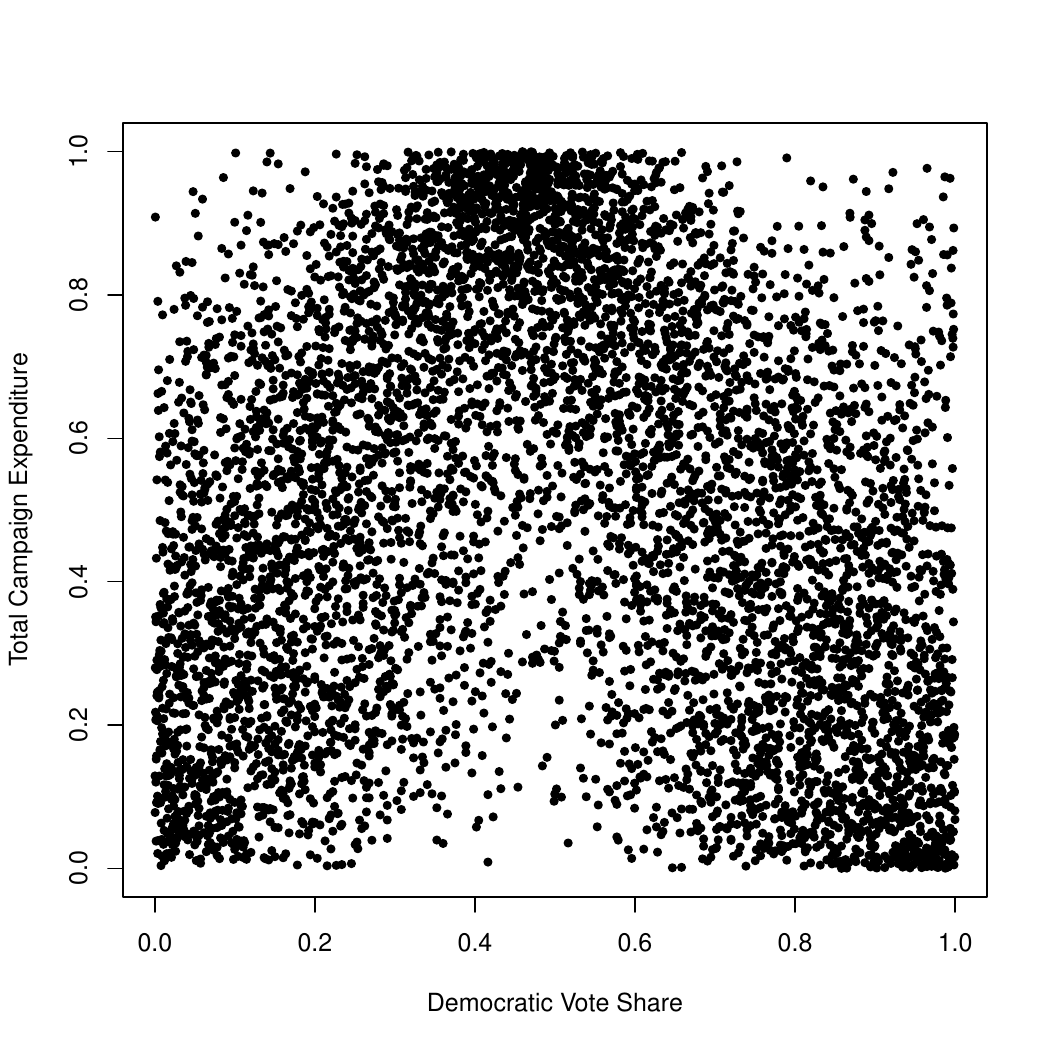}
\caption{Joint Distribution of Pseudoobservations: Vote Share and Expenditure, U.S.~House Elections, 1946-2014}
\label{election_U}
\end{figure}

As an example set of variables that follow a normal mode copula, I focus on \citeauthor{Jacobson:2015}'s (\citeyear{Jacobson:2015}) dataset about the Democratic vote share ($X_{1}$) and the total campaign expenditure ($X_{2}$) that we saw in the introduction.
In general, before analysts apply a normal mode copula to any data, they have to nail down the values of $\kappa_{1}$ and $\kappa_{2}$, which are model choice indicators rather than parameters to be estimated. In our case, a visual inspection of the scatter plot of the pseudoobservations $\hat{u}_{1, i}$ and $\hat{u}_{2, i}$ (Figure \ref{election_U}) suggests that they follow the normal mode copula with $\kappa_{1} = 2$ and $\kappa_{2} = 1$ (cf., the third panel of Figure \ref{example.plot}).
For the purpose of comparison, I also consider five conventional copulas: Ali-Mikhail-Haq (AMH), Clayton, FGM, Frank, and Gaussian. The Gaussian and FGM copulas have already been defined above (Sections \ref{sec:definition} and \ref{sec:other}). 
The other three are Archimedean copulas have generator functions 
\begin{align*}
\varphi_{\mathrm{AMH}}(u) &\equiv \log \left[ \frac{1}{u} \{ 1 - \theta (1 - u) \} \right]& \textrm{for} & \quad -1 \leq \theta \leq 1\\
\varphi_{\mathrm{Clayton}}(u) &\equiv \theta^{-1} (u^{- \theta} - 1) & \textrm{for} & \quad \theta> 0\\
\varphi_{\mathrm{Frank}}(u) &\equiv - \log \left[ \frac{1}{\exp( - \theta ) - 1} \{ \exp( - \theta u) - 1 \}\right] & \textrm{for} & \quad \theta \in \mathbb{R}.\\
\end{align*}
Unless these copulas are reduced to the product copula (where variables are independent of each other), their linear correlation is not equal to zero.

\begin{center}
[Figure 3 about here]
\end{center}

Table \ref{result} reports the results of the analysis.\footnote{I use the statistical computational environment \texttt{R} \citep{R}. For estimation of $\theta$ of the five conventional copulas, I use the \texttt{copula} library \citep{copula_library}. For estimation of $\theta$ of the normal mode copula and calculation of CvMC, AIC, and CIC of all six copulas, I write my own script, which is available at Harvard Dataverse upon acceptance. }
The first column presents $\hat{\theta}$.
In the second column, we find CvMC.
The normal mode copula achieves the lowest CvMC and thus is the best in terms of goodness-of-fit.
The third and fourth columns show AIC and CIC, respectively.
I multiply CIC by $-2N$ so that AIC and CIC are comparable. No doubt, in terms of either criterion, the normal mode copula is most preferred.
Taken together, as I guess at first sight, the normal mode copula with $\kappa_{1} = 2$ and $\kappa_{2} = 1$ is the best modeling approach to the data among the six copulas that I study.

\begin{center}
[Table 1 about here]
\end{center}

\section{Discussion}

This study focuses on an overlooked family of copulas, normal mode copulas, and characterizes its properties (Section \ref{sec:Properties}). The most important feature is that normal mode copulas can model a joint distribution where multiple variables are not linearly correlated with each other but are still dependent. Most measures of association are equal to zero when a mode number $\kappa_{d}$ is even (Proposition \ref{Measures.of.Association}). Relatedly, the dependence structure that a normal mode copula represents is not monotonic unless $\kappa_{1} = \kappa_{2} = 1$ (Propositions \ref{monotonicity} and \ref{Dependence.Ordering}). Furthermore, I apply a normal mode copula to a dataset of vote share and campaign expenditure to show that the normal mode copula achieves higher performance than other conventional copulas (Section \ref{sec:Application}).

There remain future research agenda. Scholars can apply normal mode copulas to various data to find interesting dependence structures. In my companion paper, \citet{Fukumoto:2023} analyzes the relationship between government formation and duration by taking into account some covariates to model parameters of marginal distributions ($F_{d}$). Multivariate normal mode copulas (Section \ref{sec:definition_multivariate}) as well as extended normal mode copulas (Section \ref{sec:other}) are worth exploring. 
I hope normal mode copulas help us to deeply understand mutually dependent variables.

\section*{Aknowledgments}

Earlier versions of this paper were presented on November 5, 2013, June 23, 2015, and September 16, 2022, at the Institute of Statistical Mathematics, Tokyo, Japan, and on January 6–7, 2023, at the 10th Asian Political Methodology Meeting, National University of Singapore. I thank Toshikazu Kitano and Dean Knox for their helpful comments.

\section*{Declaration of Interest Statement}

The author reports there are no competing interests to declare.

\section*{Funding Details}

This work was supported by the Japan Society for the Promotion of Science [grant number KAKENHI JP19K21683].

\section*{Data Availability Statement} 

The replication materials will be hosted on the Harvard Dataverse site (\url{https://dataverse. harvard.edu/}) upon acceptance.

\bibliographystyle{chicago}
\bibliography{ref}

\newpage

\begin{table}[ht]
\centering
\caption{Results for Applying Six Copulas to the Data of Vote Share and Expenditure}
\begin{tabular}{lrrrr}
 \hline
Copula & $\hat{\theta}$ & CvMC & AIC & $-2N \times$ CIC \\ 
 \hline
Normal Mode & 1.00 & 1.12 & $-$1842.22 & $-$1847.02 \\ 
 AMH & $-$0.87 & 6.06 & $-$190.61 & $-$189.53 \\ 
 Clayton & 0.02 & 6.50 & $-$1.31 & $-$1.28 \\ 
 FGM & $-$0.36 & 5.51 & $-$89.43 & $-$89.43 \\ 
 Frank & $-$0.70 & 5.51 & $-$86.84 & $-$86.92 \\ 
 Gaussian & $-$0.11 & 5.54 & $-$83.25 & $-$83.80 \\ 
 \hline
\end{tabular}
\label{result}
\end{table}

\newpage

\newpage

\newpage

\newpage

\section*{Figure Captions }

\begin{itemize}
\item Figure 1: Joint Distribution of Vote Share and Expenditure, U.S.~House Elections, 1946-2014
\item Figure 2: Example Plots of Normal Mode Copulas
\item Figure 3: Joint Distribution of Pseudoobservations: Vote Share and Expenditure, U.S.~House Elections, 1946-2014
\end{itemize}

\end{document}